\documentclass[twocolumn,iop]{openjournal}

\usepackage[british]{babel}

\usepackage{mathtools}
\usepackage{natbib}
\usepackage[colorlinks,allcolors=blue]{hyperref}

\newcommand{\D}{\mathrm{d}}
\newcommand{\E}{\mathrm{e}}

\begin{document}

\journalinfo{The Open Journal of Astrophysics}
\submitted{submitted 1 April 2020; accepted 18 June 2020}

\shorttitle{Source Distributions in Efficiency Space}
\shortauthors{Tessore \& Harrison}

\title{Source Distributions of Cosmic Shear Surveys in Efficiency Space}

\author{Nicolas Tessore$^{\star1}$}
\author{Ian Harrison$^{\dagger1,2}$}

\affiliation{$^1$ Jodrell Bank Centre for Astrophysics, Department of Physics \& Astronomy, The University of Manchester, Manchester M13 9PL, UK}
\affiliation{$^2$ Department of Physics, University of Oxford, Denys Wilkinson Building, Keble Road, Oxford, OX1 3RH, UK}

\thanks{$^\star$ E-mail: \nolinkurl{nicolas.tessore@manchester.ac.uk}}
\thanks{$^\dagger$ E-mail: \nolinkurl{ian.harrison-2@manchester.ac.uk}}

\begin{abstract}
We show that the lensing efficiency of cosmic shear generically has a simple shape, even in the case of a tomographic survey with badly behaved photometric redshifts.
We argue that source distributions for cosmic shear can therefore be more effectively parametrised in ``efficiency space''.
Using realistic simulations, we find that the true lensing efficiency of a current cosmic shear survey without disconnected outliers in the redshift distributions can be described to per cent accuracy with only two parameters, and the approach straightforwardly generalises to other parametric forms and surveys.
The cosmic shear signal is thus largely insensitive to the details of the source distributions, and the features that matter can be summarised by a small number of suitable efficiency parameters.

For the simulated survey, we show that prior knowledge at the~10\% level, which is attainable e.g.\@ from photometric redshifts, is enough to marginalise over the efficiency parameters without severely affecting the constraints on the cosmology parameters~$\Omega_m$ and~$\sigma_8$.
\end{abstract}


\maketitle

\section{Introduction}

Measurements of cosmic shear obtained via the weak lensing effect on individual galaxy shapes are one of the best available probes of the late Universe where Dark Energy dominates.
The large numbers of galaxies necessary to reduce the statistical noise on cosmic shear two-point functions requires that current \citep{2018PhRvD..98d3528T,2019PASJ...71...43H,2020A&A...633A..69H} and future \citep{2018arXiv180901669T,2018LRR....21....2A} surveys rely on photometric detections of sources only.
This means that the second crucial piece of information necessary for cosmic shear cosmology -- distances to the galaxies for which shapes are measured -- typically comes with the large uncertainties inherent in photometric redshifts.
Large amounts of effort are expended on how to increase the precision and accuracy of such uncertain redshift estimates \citep[see e.g.][for a review of a number of methods]{2020arXiv200103621S}.

The fiducial approach for current surveys is to parametrise uncertainty on the potential shift of the mean of the redshift distribution~$n(z - \Delta z)$, with~$\Delta z$ in each tomographic bin marginalised over with a Gaussian prior.
In addition to this statistical error, the systematic uncertainty stemming from the methodological differences in how the initial~$n(z)$ is formed has also been argued to dominate over the statistical uncertainty of current surveys \citep{2020A&A...638L...1J}.
A well-motivated and principled way of accounting for a much wider range of statistical uncertainties than simply the shift in mean involves marginalising over the heights of histogram bins for the number count of weak lensing source galaxies as a function of redshift \citep[e.g.][]{2016MNRAS.460.4258L,2019MNRAS.483.2801S, 2019arXiv191007127A}, but this necessarily creates a large number of new nuisance parameters which cannot feasibly be included in a typical analysis.

\begin{figure}%
\centering%
\includegraphics[scale=0.8]{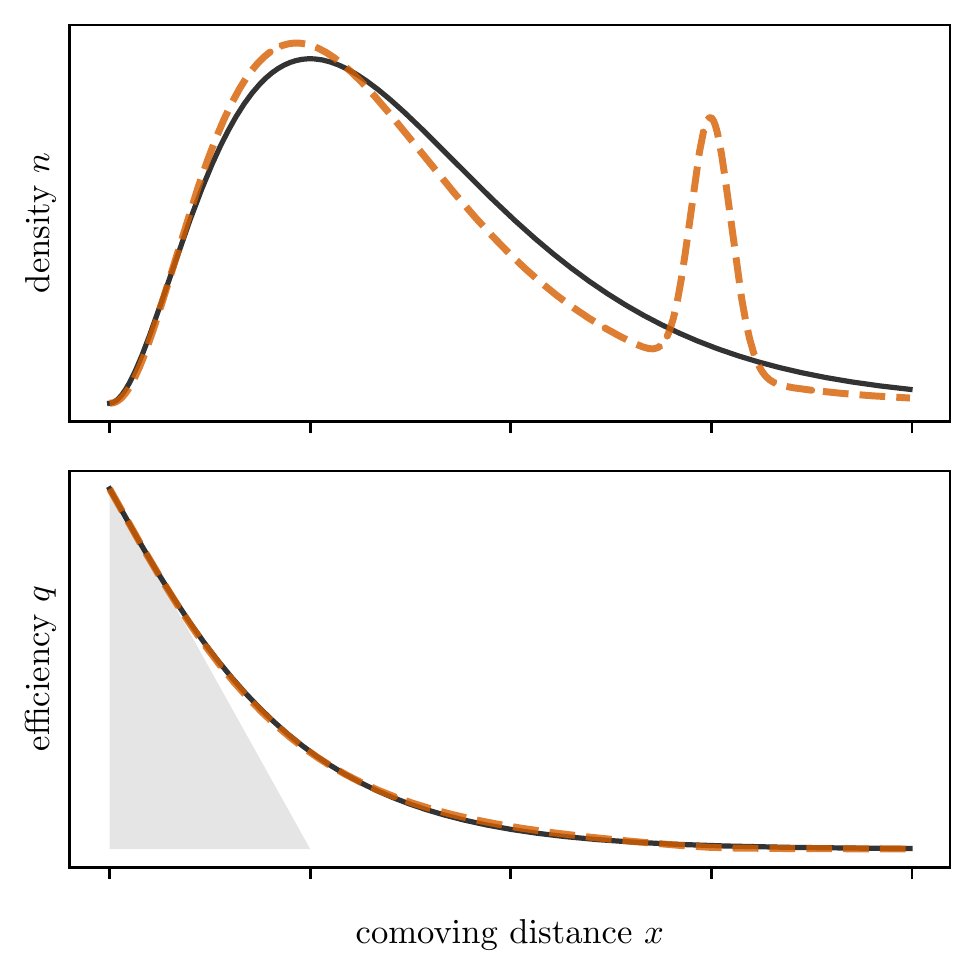}%
\caption{%
Distinct source distance distributions (\emph{top}) with visually (and cosmologically, see Section~\ref{sec:cosmology}) similar lensing efficiencies~(\emph{bottom}).
The total area under the curve~$q(x)$ is $\mu/2$, whereas it would be the shaded area $\eta^{-1}/2$ for a curve with fixed slope.
}%
\label{fig:intro}%
\end{figure}

In this paper, we argue that attention should instead be focused directly on the lensing efficiency~$q$ of the source distribution, and that this is where the constraining power of data can be most effectively expended.
The argument stems from the fact that the source distribution, in the form of the source density~$n(x)$ per comoving distance~$x$, only enters the cosmic shear signal through the lensing efficiency (for a good short summary, see \citealp{2017JCAP...05..014L}),
\begin{equation}
    q(x)
    = \int_{x}^{\infty} \! \frac{t - x}{t} \, n(t) \, \D{t} \;. \label{eq:q}
\end{equation}
However, this integral operator smooths out almost all details of the source distribution, so much so that even sharp features in~$n(x)$ can have no appreciable effect on $q(x)$, as shown in Fig.~\ref{fig:intro}.

This smoothing is a generic feature of the lensing efficiency, which can be understood as follows.
Instead of the integral~\eqref{eq:q}, the lensing efficiency is equivalently characterised by a second-order differential equation, obtained by differentiating twice,
\begin{equation}
    q''(x) = \frac{n(x)}{x} \;,
\end{equation}
with initial value $q(0) = 1$ due to the normalisation of~$n(x)$, and initial slope
\begin{equation}
    q'(0)
    = -\int_{0}^{\infty} \! \frac{n(x)}{x} \, \D{x}
    = -\eta
    \label{eq:qprime}
\end{equation}
given by the mean inverse comoving distance~$\eta$.
Hence, for fixed~$\eta$, different densities~$n(x)$ only lead to different accelerations along the curve.
However, the integrated lensing efficiency
\begin{equation}
    \int_{0}^{\infty} \! q(x) \, \D{x}
    = \mu/2
    \label{eq:qint}
\end{equation}
is also constrained by the mean comoving distance~$\mu$.
For fixed initial conditions but no acceleration (i.e.\@ $q(0) = 1$, $q'(0) = -\eta$, $q''(0) = 0$), the integrated efficiency would be $\eta^{-1}/2$, which is less than $\mu/2$ by Jensen's inequality (cf.\@ Fig.~\ref{fig:intro}).
The mean~$\mu$ of the source distribution therefore describes the tail of the curve, while the inverse mean~$\eta$ describes its behaviour near the origin.
Overall, this essentially fixes the shape of~$q(x)$.

In the following, we develop this idea into a simple method for cosmic shear in efficiency space instead of redshift space.
In Section~\ref{sec:model}, we fix a parametric form for the lensing efficiency, based on the preceding argument, and show that it can describe a current cosmic shear survey to very good accuracy.
The parametric efficiency is then used in Section~\ref{sec:cosmology} for inference of the cosmological parameters.
A brief summary and conclusion is given in Section~\ref{sec:conclusion}.

\section{Parametric Lensing Efficiency}
\label{sec:model}

Since we expect the efficiency~$q(x)$ to depend little on the details of the source density~$n(x)$, we can derive a parametric form for~$q(x)$ by introducing a convenient parametric form for~$n(x)$ and computing its efficiency via the integral~\eqref{eq:q}.
In particular, we want~$n(x)$ to be positively supported, and~$q(x)$ to be of sufficiently elementary form for easy analytic and numeric evaluation.
A natural choice is thus the gamma distribution with shape parameter~$\alpha > 0$ and scale parameter~$\beta > 0$,
\begin{equation}
    n(x; \alpha, \beta)
    = \frac{\beta^{\alpha+1} \, x^\alpha \, \E^{-\beta x}}{\Gamma(\alpha+1)} \;, \label{eq:gamman}
\end{equation}
where $\Gamma(\alpha)$ is the gamma function.
The lensing efficiency is then
\begin{equation}
    q(x; \alpha, \beta)
    = Q(\alpha+1, \beta x)
    - \frac{\beta x}{\alpha} \, Q(\alpha, \beta x) \;, \label{eq:gammaq}
\end{equation}
where $Q(\alpha, x)$ is the regularised gamma function.
This makes both the density~\eqref{eq:gamman} and the efficiency~\eqref{eq:gammaq} straightforward to work with and quick to compute.

From our initial discussion, we expect the mean~$\mu$ and inverse mean~$\eta$ to be important descriptors for the shape of the lensing efficiency.
Computing both for the density~\eqref{eq:gamman},
\begin{equation}
    \mu = \frac{\alpha + 1}{\beta}
    \quad \mbox{and} \quad
    \eta = \frac{\beta}{\alpha} \;,
\end{equation}
we can readily invert the relations to parametrise the efficiency in terms of mean~$\mu$ and inverse mean~$\eta$,
\begin{equation}
    \alpha = \frac{1}{\mu\eta - 1}
    \quad \mbox{and} \quad
    \beta = \frac{\eta}{\mu\eta - 1} \;.
\end{equation}
We expect the parameters~$\mu$ and~$\eta$ to be good generic descriptors for the shape of the lensing efficiency.
The parameters~$\alpha$ and~$\beta$, on the other hand, only belong to the specific parametric form~\eqref{eq:gammaq} for the efficiency.
Positive values of~$\alpha$ and~$\beta$ always fulfil the strict constraint that $\mu\eta \ge 1$ due to Jensen's inequality.

To demonstrate that the lensing efficiency can be parametrised by~$\mu$ and~$\eta$, we apply this description to the Buzzard synthetic sky catalogue \citep{2019arXiv190102401D}, which simulates the Dark Energy Survey Year 1 (DES~Y1) observations with realistic uncertainties.
In particular, the catalogue contains both intrinsic source redshifts and photometric redshifts obtained from simulated observations with the \texttt{BPZ} algorithm \citep{2000ApJ...536..571B}, in the same way that photometric redshifts are estimated from actual DES~Y1 data \citep{2018MNRAS.478..592H}.
In total, we use three different types of redshifts from the catalogue, \emph{i)}~the intrinsic redshifts~$z_{\rm intr}$ that correspond to the comoving distance~$x$ of sources, \emph{ii)}~the photometric redshifts~$z_{\rm phot}$ used to create the photometric source distributions, obtained by a random draw from the \texttt{BPZ} $p(z)$ posterior probability, and \emph{iii)}~the photometric redshifts~$z_{\rm tomo}$ used for tomographic redshift binning, given by the mean of the \texttt{BPZ} $p(z)$ posterior probability, as in DES~Y1.

\begin{figure}%
\centering%
\includegraphics[scale=0.8]{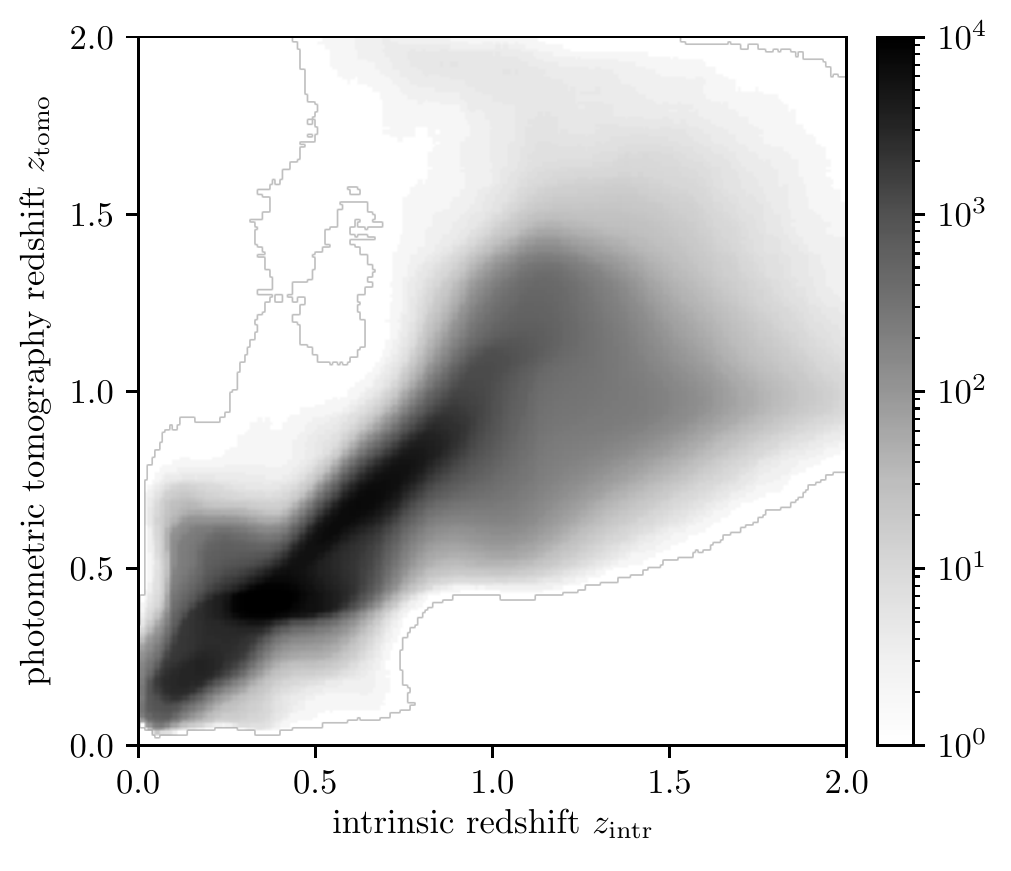}%
\caption{%
Two-dimensional distribution of intrinsic redshifts~$z_{\rm intr}$ and the photometric redshifts~$z_{\rm tomo}$ used for tomographic binning.
Source selection into tomographic bins corresponds to horizontal cuts in the plane.
Since the distribution is not purely diagonal, this creates complicated intrinsic redshift distributions when using photometric tomography.
}%
\label{fig:z_intr-z_tomo}%
\end{figure}

Fig.~\ref{fig:z_intr-z_tomo} shows the two-dimensional distribution of the intrinsic redshift~$z_{\rm intr}$ and the associated photometric redshift~$z_{\rm tomo}$ for tomographic binning.
Degeneracies in features of spectral energy distributions, such as the Lyman and Balmer breaks, combine with measurement error-induced scatter to result in some regions of~$z_{\rm intr}$ separated by~$\Delta z \sim 1$ being indistinguishable from each other.
The effect is mitigated by the use of prior distributions on $p(z)$, which in turn can be highly sensitive to selection effects and the misidentification of sources in the samples used to form the priors \citep[e.g.][]{2020arXiv200310454H}.
The resulting joint distribution of~$z_{\rm intr}$ and~$z_{\rm tomo}$ contains diffuse tails away from the diagonal, leading to tomographic source distributions in which the intrinsic redshifts may fall significantly outside the nominal bin edges.

\begin{figure}%
\centering%
\includegraphics[scale=0.8]{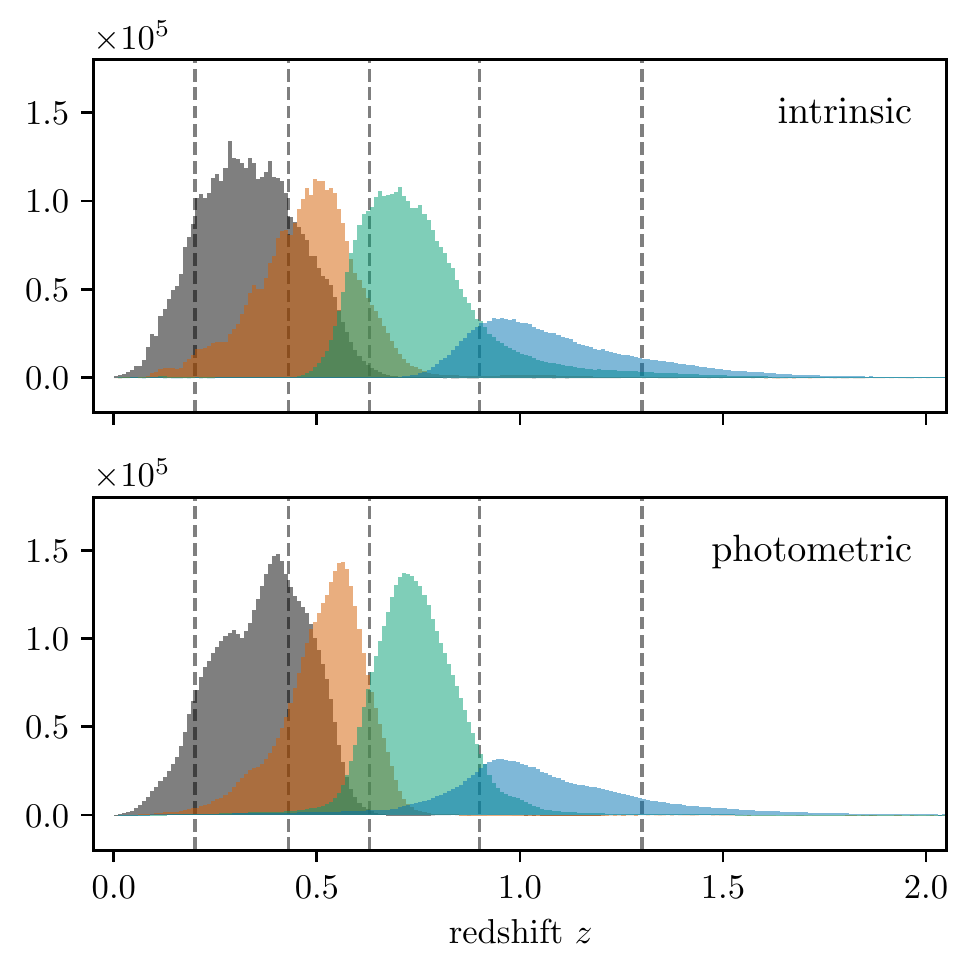}%
\caption{%
Number count histograms for intrinsic redshifts~$z_{\rm intr}$ (\emph{top}) and photometric redshifts~$z_{\rm phot}$ (\emph{bottom}) after selection of the photometric redshifts~$z_{\rm tomo}$ into nominal tomographic redshift bins (\emph{dashed lines}).
}%
\label{fig:nz-phot-tomo}%
\end{figure}

This effect of broadening and overlap of the intrinsic redshift distributions is clearly visible in Fig.~\ref{fig:nz-phot-tomo}, which shows the tomographic source distributions of the catalogue.
Here and in the following, we always assume the DES~Y1 tomographic redshift bins with bin edges of 0.2, 0.43, 0.63, 0.90, and 1.30.
While a point estimate~$z_{\rm tomo}$ is used for binning, the redshift sample~$z_{\rm phot}$ from the full posterior is used to create the shown photometric redshift distributions.
The stacking procedure often leads to biases when using photometric redshift distributions \citep{2020arXiv200103621S}, which we will shortly see via the lensing efficiency.

\begin{figure}%
\centering%
\includegraphics[scale=0.8]{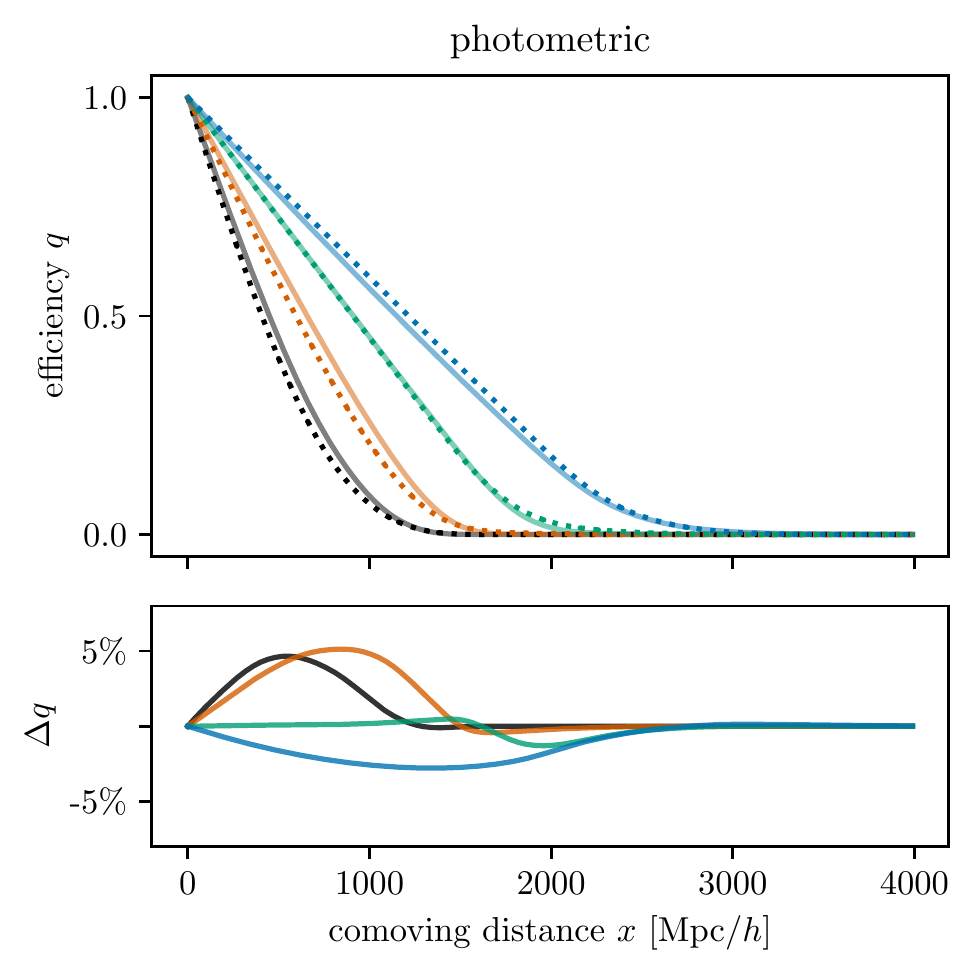}%
\caption{%
\emph{Top:}
Tomographic lensing efficiency~$q(x)$ from photometric redshifts (\emph{solid}) and intrinsic redshifts (\emph{dotted}).
\emph{Bottom:}
Absolute difference~$\Delta q(x)$ between photometric and intrinsic lensing efficiencies.
}%
\label{fig:qx-phot-tomo}%
\end{figure}

To convert redshifts to comoving distances, we use the flat $\Lambda$CDM cosmology of the Buzzard simulations with~$\Omega_m = 0.286$ \citep{2019arXiv190102401D}.
We work in units of Mpc/$h$ to remove the dependency of the comoving distances on the Hubble parameter~$h$.
The exact lensing efficiency~$q_{\rm samp}$ for a sample of sources at distances~$x_1, x_2, \ldots$ with weights~$w_1, w_2, \ldots$ can be computed as the weighted average
\begin{equation}
    q_{\rm samp}(x)
    = \frac{\sum_{x_i > x} w_i \, (x_i - x)/x_i}{\sum_{i} w_i} \;.
\end{equation}
The exact computation is free from a choice of binning for the number count histograms, which are only used for illustration.
The resulting lensing efficiencies for photometric and intrinsic redshifts are shown in Fig.~\ref{fig:qx-phot-tomo}.
The absolute error~$\Delta q$ of the photometric efficiency is at the~5\% level for the two lower tomographic bins.

\begin{figure}%
\centering%
\includegraphics[scale=0.8]{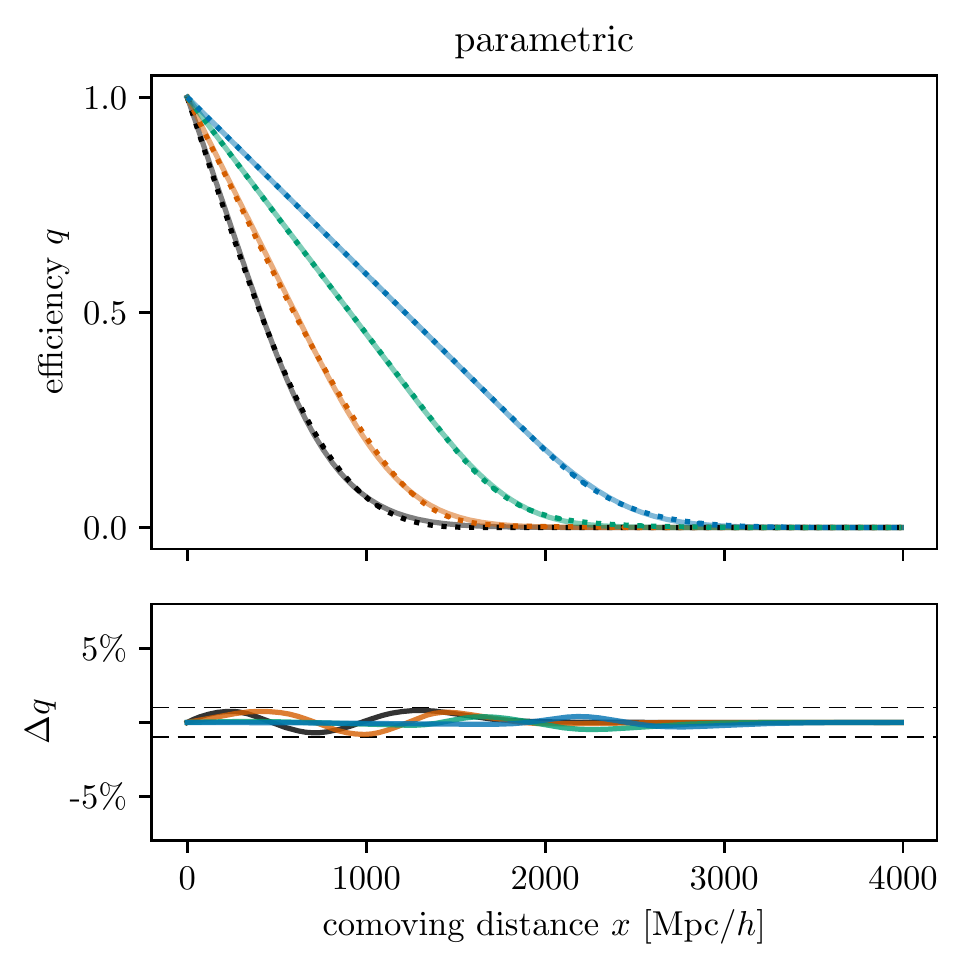}%
\caption{%
\emph{Top:}
Parametric (\emph{solid}) and intrinsic (\emph{dotted}) lensing efficiencies~$q(x)$.
\emph{Bottom:}
Absolute difference~$\Delta q(x)$ between parametric and intrinsic efficiencies.
All curves lie inside the~1\% band (\emph{dashed}).
}%
\label{fig:qx-para-tomo}%
\end{figure}

As expected, the lensing efficiencies are of the characteristic simple shape even for a realistic galaxy catalogue and photometric source selection into tomographic bins.
To show how closely the intrinsic efficiencies match the parametric form~\eqref{eq:gammaq}, we perform a continuous least squares fit by minimising the integrated square error,
\begin{equation}
    \operatorname*{arg\,min}_{\alpha, \beta} \int_{0}^{\infty} \! \bigl[q(x; \alpha, \beta) - q_{\rm samp}(x)\bigr]^2 \, \D{x} \;.
\end{equation}
The resulting parametric efficiencies are shown in Fig.~\ref{fig:qx-para-tomo}.
We find that our simple parametrisation reproduces the intrinsic efficiencies to per cent accuracy across the entire distance range and all tomographic bins and, in Section~\ref{sec:cosmology}, that this accuracy is sufficient for useful cosmological constraints.
The best-fit efficiency parameters~$\mu$ and~$\eta$ are given in Table~\ref{tab:fit}.
They are in good agreement with the parameters obtained directly from the intrinsic source distributions.

\begin{table}%
\centering%
\caption{%
Mean~$\mu$~[Gpc/$h$] and inverse mean~$\eta$~[$h$/Gpc] for the intrinsic and photometric source distributions, and the parametric efficiency.
}%
\label{tab:fit}%
\begin{tabular}{ccccccc}
\hline
& \multicolumn{2}{c}{intrinsic} & \multicolumn{2}{c}{photometric} & \multicolumn{2}{c}{parametric} \\
bin & $\mu$ & $\eta$ & $\mu$ & $\eta$ & $\mu$ & $\eta$ \\
\hline
1 & 0.920 & 1.301 & 0.989 & 1.156 & 0.929 & 1.235 \\
2 & 1.273 & 0.855 & 1.365 & 0.769 & 1.278 & 0.832 \\
3 & 1.865 & 0.554 & 1.856 & 0.552 & 1.860 & 0.551 \\
4 & 2.476 & 0.412 & 2.387 & 0.449 & 2.473 & 0.412 \\
\hline
\end{tabular}%
\end{table}

\begin{figure}%
\centering%
\includegraphics[scale=0.8]{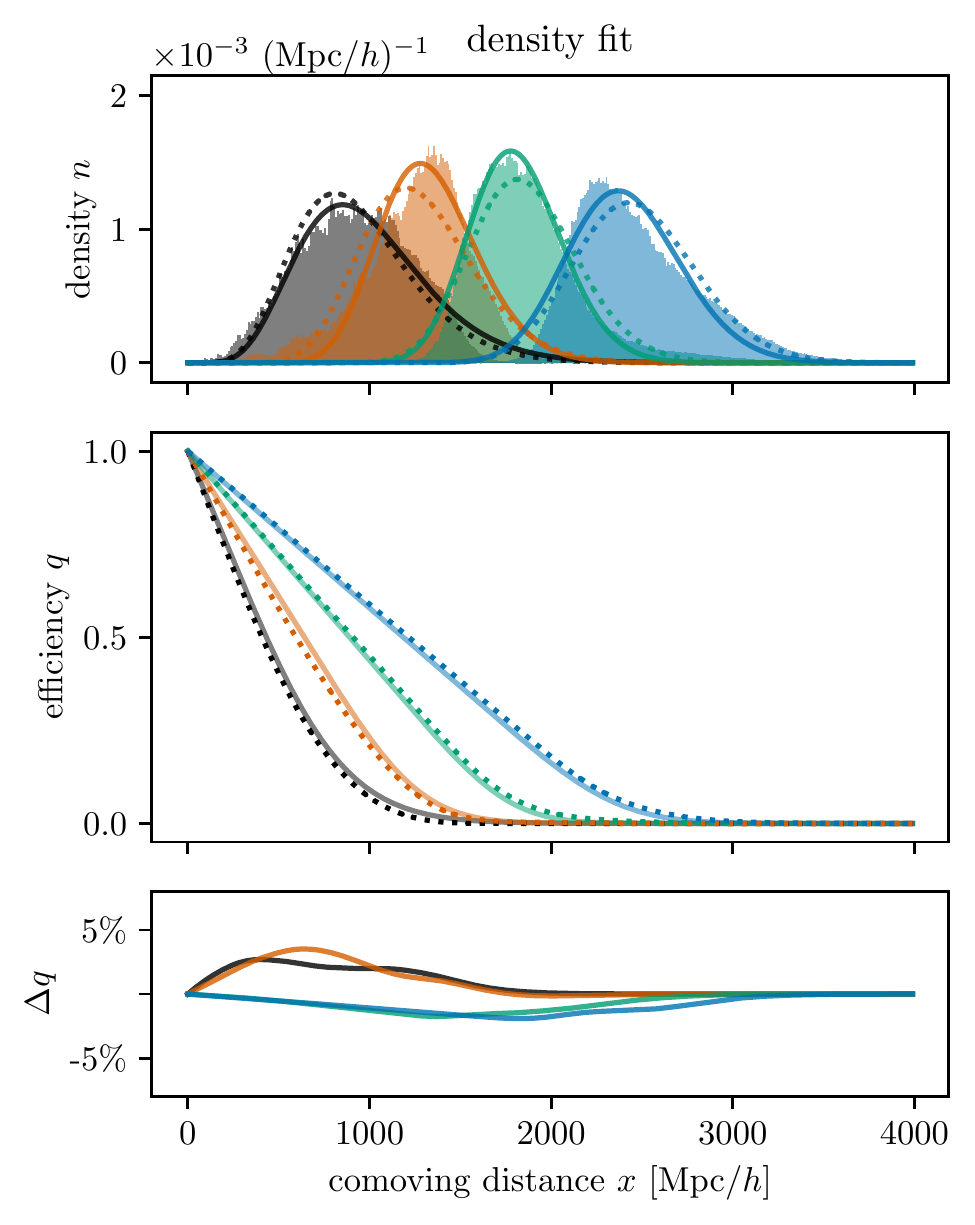}%
\caption{%
\emph{Top:}
Best-fit densities (\emph{solid}) and densities for the parametric efficiencies (\emph{dotted}).
\emph{Middle:}
Lensing efficiencies for the best-fit densities (\emph{solid}) and intrinsic efficiencies (\emph{dotted}).
\emph{Bottom:}
Absolute difference $\Delta q(x)$ between efficiencies for the best-fit densities and intrinsic efficiencies.
}%
\label{fig:nx-qx-nfit-tomo}%
\end{figure}

Overall, we find that the efficiency parameters~$\mu$ and~$\eta$ suffice to describe a DES~Y1-like cosmic shear survey.
We can furthermore recover the lensing efficiency through a simple parametric form~\eqref{eq:gammaq}.
This is not merely due to similarities between the assumed parametric form~\eqref{eq:gamman} of the source distributions and the catalogue:
Fig.~\ref{fig:nx-qx-nfit-tomo} shows that a direct least squares fit of the density~$n(x)$ to the intrinsic distributions produces a better match but yields significantly degraded efficiencies that are accurate only to a level comparable to photometric redshifts.
The fitting in density space could be improved through a loss function that is specifically chosen to match the cosmological information; this happens naturally in efficiency space.
The parametric form~\eqref{eq:gammaq} for the efficiency is nevertheless only a convenient choice, while the generic observation is that the lensing efficiency is almost featureless and easily parametrised by a suitable density function.
This could be e.g.\@ a Gaussian, suitably clipped to the positive reals, or a generalised gamma distribution, in which case the agreement with the intrinsic efficiencies of the Buzzard catalogue improves by more than a factor of two.

\begin{figure}%
\centering%
\includegraphics[scale=0.8]{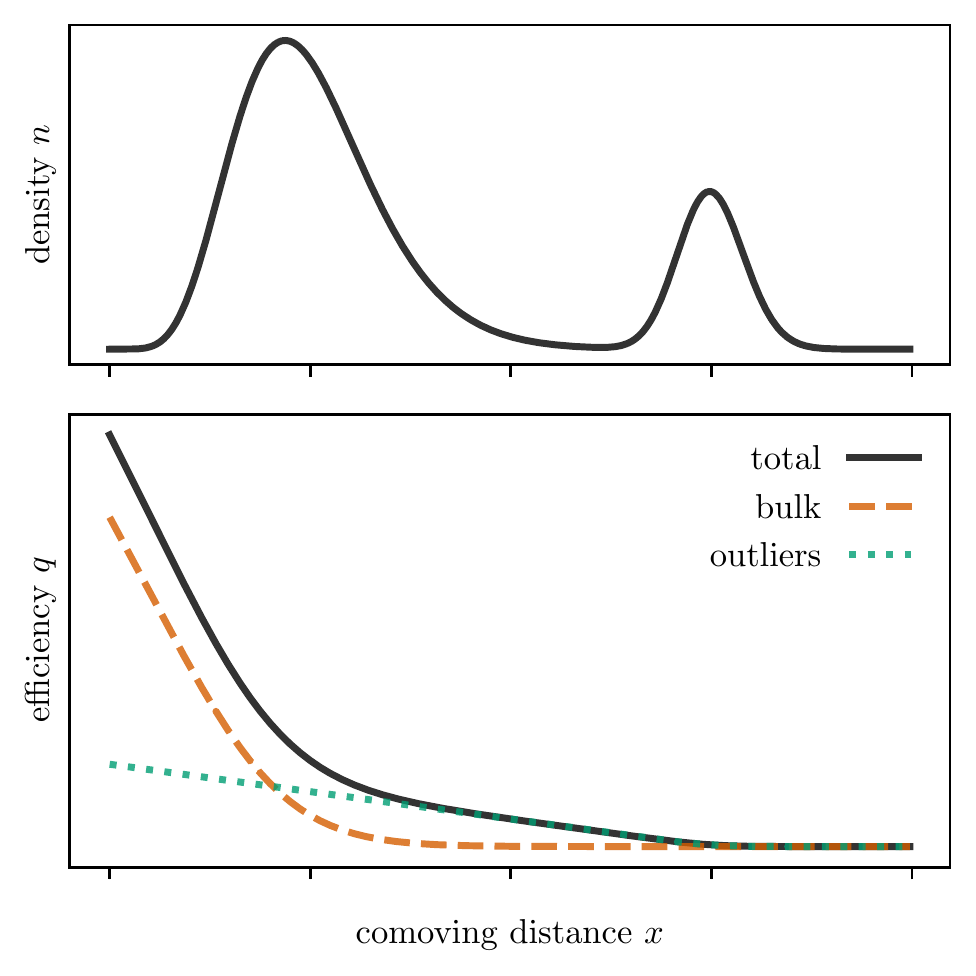}%
\caption{%
The effect of disjoint outlier populations on the lensing efficiency.
The total lensing efficiency (\emph{solid}) has two distinct slopes corresponding to the lensing efficiencies of the bulk (\emph{dashed}) and the outlier (\emph{dotted}) population.
}%
\label{fig:outliers}%
\end{figure}

Parametric lensing efficiencies obtained in this way, i.e.\@ by integrating a chosen~$n(x)$, describe catalogues with outliers that are connected to the bulk of the tomographic source distributions.
It is well known that outliers can strongly bias the cosmic shear signal \citep{2007MNRAS.381.1018A}, and this effect can be seen in terms of the efficiency parameters, since even a small fraction of outliers can severely impact the mean and inverse mean.
Such outliers are expected to comprise populations of physically related galaxies, which appear similar in photometric observations of low spectral resolution, but are actually separated in redshift.
If this separation is large enough, tomographic selection by photometric redshifts may result in multiple disjoint populations in the source distributions.
The total, composite distribution of such a sample has a lensing efficiency with a modified shape that is the superposition of simple shapes for the individual populations.
This is shown schematically in Fig.~\ref{fig:outliers}, in which the differing contributions to the total efficiency from two disjoint populations can be seen.
For surveys where disjoint outlier populations are significant, it is therefore possible to use a mixture of parametric efficiencies to describe each component individually.
However, as seen above, this is not necessary for the Buzzard catalogue, where the total efficiency is well described by a single component.

\section{Cosmological Parameter Inference}
\label{sec:cosmology}

We now use the parametric efficiencies to infer the cosmological parameters~$\Omega_m$ and~$\sigma_8$ from measurements of cosmic shear.

In a first step, we simulate a shear-only two-point function data vector, mimicking the corresponding DES~Y1 data product \citep{2017arXiv170609359K}, using the Buzzard catalogue's intrinsic redshifts and cosmology parameters $\Omega_m = 0.286$, $\sigma_8 = 0.82$, and $h = 0.70$ \citep{2019arXiv190102401D}.
We use a \texttt{CosmoSIS} \citep[\texttt{v1.6}]{2015A&C....12...45Z} pipeline that runs \texttt{CAMB} \citep{2012JCAP...04..027H,2000ApJ...538..473L} and \texttt{Halofit} \citep{2012ApJ...761..152T,2003MNRAS.341.1311S} to obtain the matter power spectrum, which is projected using Limber's approximation to produce the tomographic shear power spectra \citep[for details, see e.g.\@][]{2018PhRvD..98d3526A}, and transformed to the shear two-point functions \citep[using the method of][]{2009A&A...497..677K}.
These two-point functions are then stored as our simulated data vector, together with a Gaussian covariance matrix matching the effective number densities and shape noise of DES~Y1 \citep{2018PhRvD..98d3528T}.

The second step is the analysis of the synthetic data vector using the same pipeline, but with~$\Omega_m$ and~$\sigma_8$ left as free parameters to be constrained by the data.
The Hubble parameter~$h$ is fixed to the true value; this does not affect the results since cosmic shear-only results are highly insensitive to the value of $h$, and the efficiency parameters~$\mu$ and $\eta$ are given in Mpc/$h$ in order to not artificially break this degeneracy.
The analysis is performed in redshift space using \emph{i)}~the intrinsic redshifts, and \emph{ii)}~in efficiency space using the parametric efficiencies fixed to the best-fit values.

\begin{figure}%
\centering%
\includegraphics[scale=0.8]{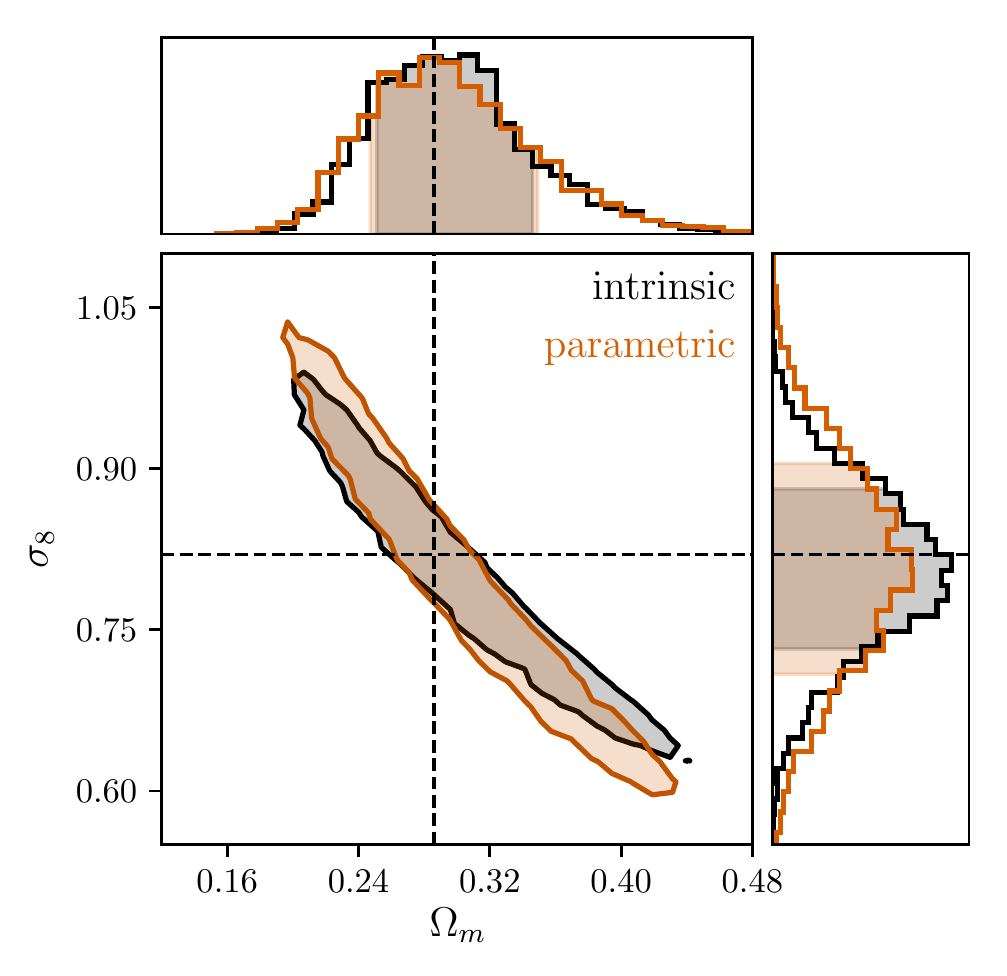}%
\caption{%
Constraints on the cosmology parameters~$\Omega_m$ and $\sigma_8$.
Shown are the two sigma joint contours and marginal distributions from the intrinsic redshifts (\emph{black}), and using the best-fit parametric efficiencies (\emph{orange}).
The degeneracy changes because parametric efficiencies do not use the cosmology to convert redshifts to comoving distances.
}%
\label{fig:contours}%
\end{figure}%

The resulting two sigma contours for $\Omega_m$ and $\sigma_8$ are shown in Fig.~\ref{fig:contours}.
Using a parametric efficiency affects the degeneracy of the cosmological parameters, because the cosmology is not used to convert the redshift distributions to comoving distance.
This leads to a tilt in the contours between the efficiency space and redshift space analysis.
The marginal distributions are largely unaffected, with a mildly wider marginal posterior for~$\sigma_8$, and no change in the marginal posterior for~$\Omega_m$.
Neither cosmological parameter is biased by the parametric efficiency with respect to the intrinsic redshifts.

The above shows that for known source distributions, working in efficiency space yields roughly the same results as working in redshift space.
The real advantage of a parametric efficiency lies in the opposite direction, where it becomes possible to do cosmic shear cosmology, in the extreme case, without any information about the source distributions.
Leaving the efficiency parameters as free parameters allows sampling of the posterior while exploring all possible source distributions that are covered by the efficiency model (such as, in the case we are considering, those without disjoint outlier populations).
Any information about the sources that is available can then be incorporated into the analysis through a prior on the efficiency parameters, as appropriate for a Bayesian analysis.
In the efficiency space approach, the data entering in the observed shear two-point function could contain now redshift information directly, with this entering only through the prior, which could be informed by an external data set, which may be deeper or with more available photometric (or spectroscopic) bands. Whilst this is typical of current methods working in density space, using this prior in efficiency space reduces loss of information and relatively uninformative priors (which can safeguard against biases) are possible.

To understand how imperfect knowledge about the efficiency parameters affects the cosmology, we repeat the same analysis, but instead of fixing the efficiency parameters to the best-fit values, we equip~$\mu$ and~$\eta^{-1}$ with a uniform prior of varying width.
We use the reciprocal~$\eta^{-1}$ under the principle that a distance, not an inverse distance, should be uniformly distributed.
We set the uniform prior range of all efficiency parameters to a fraction~$f$ of their respective true value,
\begin{equation}
    (1-f) \, P_i^{\rm true} \le P_i \le (1+f) \, P_i^{\rm true} \;,
\end{equation}
where the parameter~$P$ is either $\mu$ or $\eta^{-1}$, and~$i$ is the tomographic bin index.
In every instance, the efficiency parameter space is also naturally bounded by the strict condition that $\mu_i\eta_i \ge 1$, which the sampling takes into account.
We sample the posterior distribution at uniform prior widths of 2\%, 5\%, 10\%, 20\%, and 100\%, so that we can obtain the posteriors at intermediate widths by resampling without unduly reducing the number of effective samples.

\begin{figure}%
\centering%
\includegraphics[scale=0.8]{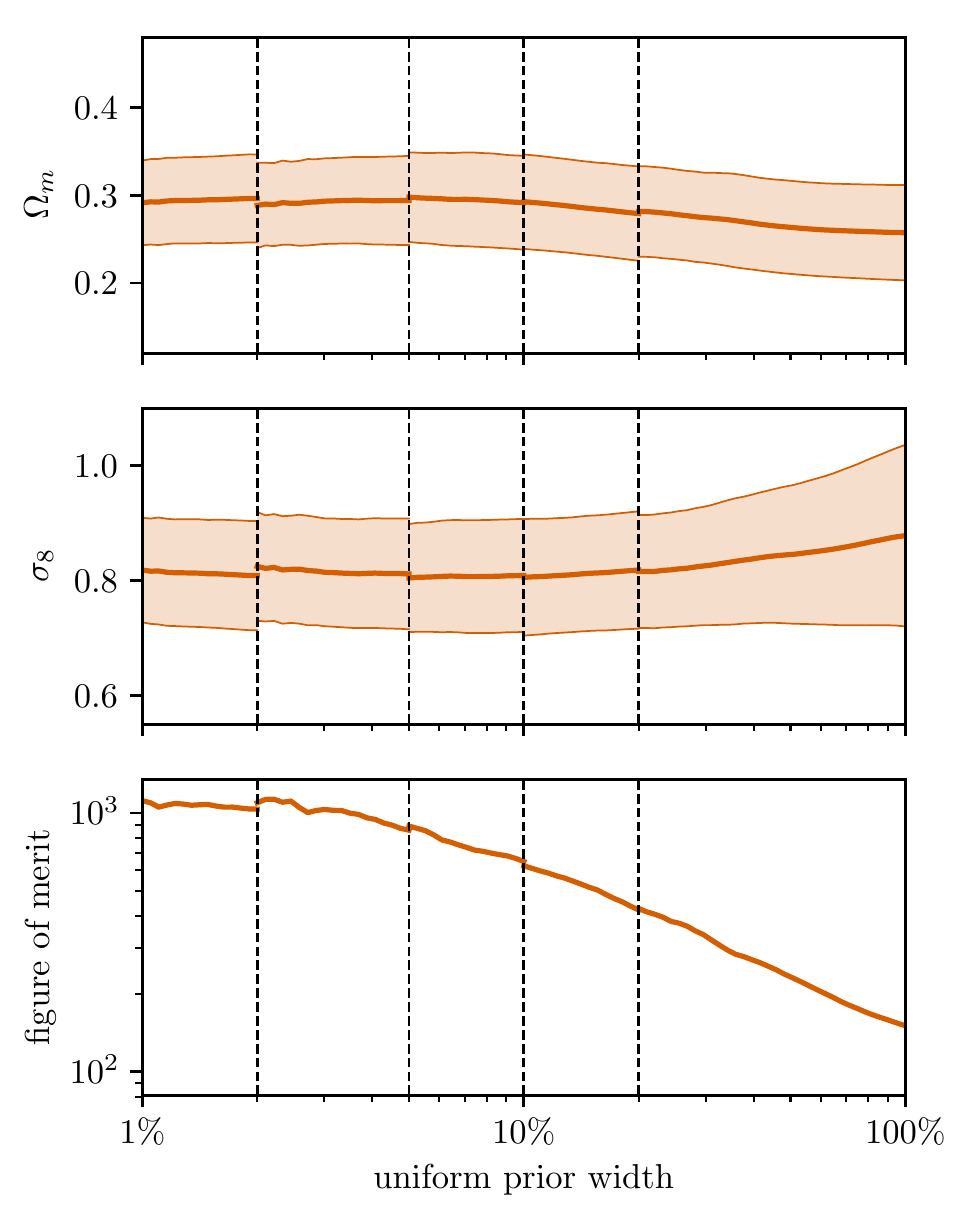}%
\caption{%
Mean and standard deviation of the marginal posterior distribution for~$\Omega_m$ and~$\sigma_8$ as functions of the relative width of uniform priors on the efficiency parameters.
The figure of merit quantifies the constraining power of the joint distribution.
Dashed lines indicate a change in the underlying chains.
}%
\label{fig:priors}%
\end{figure}

The resulting cosmological constraints are shown in Fig.~\ref{fig:priors}.
The mean and standard deviation of the marginal distributions of~$\Omega_m$ and~$\sigma_8$ are relatively stable to uniform prior widths of~${\sim} 10\%$, from which point on the marginal means develop a slight bias, which grows up to~${\sim} 1$ standard deviation at $100\%$ prior width.
Interestingly, the width of the $\Omega_m$ constraint remains constant over the entire range, while the constraint for $\sigma_8$ widens by a factor of~${\sim} 1.5$.
To quantify how broadening the efficiency parameter priors affects the joint constraining power for the cosmological parameters, we define a figure of merit as the inverse of the area of the covariance ellipse \citep{2006astro.ph..9591A},
\begin{equation}
    \mathrm{FoM}
    = \bigl[\det \operatorname{Cov}(\Omega_m, \sigma_8)\bigr]^{-\frac{1}{2}} \;.
\end{equation}
We have checked this figure of merit against a subset of $\Omega_m$-$\sigma_8$ contour areas and found them to very closely agree, with the same behaviour as the prior width is increased, implying that parameter degeneracies (which can be strong) are not a problem for our figure of merit as defined for this experimental configuration.

Using the figure of merit, we find again that a uniform prior width below ${\sim} 10\%$ does not severely affect the posterior. For larger prior widths, the figure of merit falls by a factor ${\sim} 10$ at $100\%$.
This, together with the results for the marginal distributions, implies that the joint contours for~$\Omega_m$ and~$\sigma_8$ become rounder as prior information about the source distributions is removed, but not significantly wider along either marginal axis.

This qualitative analysis demonstrates the feasibility of obtaining real cosmology constraints with the efficiency space approach when the efficiency parameters are not perfectly known.
With a look at Table~\ref{tab:fit}, we find that the photometric values of $\mu$ and $\eta$ lie within the ${\sim} 10\%$ range at which the analysis starts to be adversely affected by the lack of information about the sources.
Furthermore, uniform priors for~$\mu$ and~$\eta$ could readily be improved through targeted modelling and/or higher-dimensional and hierarchical sampling methods for the distance distributions.
We also emphasise that these results are meant to show the sensitivity of cosmological parameters to the efficiency parameters, and should not be taken as a proposal for, or compared to, full analyses.
Our message in this paper is to demonstrate the principle of modelling distance distributions in efficiency space.


\section{Conclusion}
\label{sec:conclusion}

We have shown that the distribution of distances to sources in a cosmic shear analysis can be modelled directly in the space of the lensing efficiency, rather than in the space of the source redshifts.
We have argued that the approach is motivated by the form of the lensing efficiency transformation, which smooths out many features of source redshift distributions.
This means that expending parameters on modelling such features is unnecessary when the goal of the analysis is cosmic shear cosmology.
By modelling the source distance distribution in efficiency space, we are modelling only the information that is necessary for cosmology.
The behaviour of the lensing efficiency in~\eqref{eq:qprime} and~\eqref{eq:qint} suggests that only the parameters~$\eta$ and~$\mu$ are necessary to describe the weak lensing action of a single population of sources, and Fig.~\ref{fig:outliers} shows how this readily and simply extends to outlier populations.

We have chosen a parametrised form for the lensing efficiency~\eqref{eq:gammaq} and shown in Fig.~\ref{fig:qx-para-tomo} that for a representative cosmic shear survey (DES~Y1 as modelled by the Buzzard simulation) two free parameters per tomographic bin allow us to model the true efficiency to within~1\% accuracy.
The further analysis of Section~\ref{sec:cosmology} then shows the effect of this on the constraints for~$\Omega_m$ and~$\sigma_8$, the cosmological parameters best constrained by cosmic shear surveys, where we find little loss in constraining power even when marginalising over efficiency parameters which are uncertain at the level of up to~10\%.
These priors on efficiency distribution parameters could, but need not, be informed by other methods (such as Hierarchical Bayesian Methods) which explicitly draw samples from the redshift distribution.

The most informative cosmological analysis from photometric surveys are typically done by combining weak lensing shear with galaxy cluster and galaxy-galaxy lensing \citep[e.g.][]{2018PhRvD..98d3526A, 2018MNRAS.476.4662V}.
In the case where the ``source'' sample of galaxies (from which the shear is measured) and the ``lens'' sample (around which shear is measured, and the clustering of which is measured) are disjoint samples of objects, the natural (and likely optimal) approach is to describe the lenses as usual in density space, and the sources in efficiency space.
For cases where this separation is not complete, or where details of the density distribution become important for cosmic shear \citep[such as in the case for Intrinsic Alignments of galaxies, e.g.][]{2015SSRv..193....1J} the optimal trade off between bias and variance induced by different approaches will depend on the details of the source distributions and the experiment.
One could either model the distribution for all samples in efficiency space (which is optimal for shear), and subsequently cope with the biases created in the density space part of the analysis (i.e.\@ lenses and Intrinsic Alignments), or vice-versa.
We leave the exploration of this to future work.

More sophisticated methods are necessary for cosmological parameter estimation in real data, but here we have argued modelling of that data can be done most parsimoniously by modelling the lensing efficiency directly, rather than the redshift number density distribution.

\pagebreak[0]

\section*{Acknowledgements}

We thank J.~DeRose and R.~H.~Wechsler for providing us with advance access to the Buzzard DES~Y1 catalogue, and for helping with its use.
We also thank S.~Bridle, J.~P.~Cordero, R.~P.~Rollins, and A.~Amara for their insights and many fruitful discussions.

The preparation of this manuscript was made possible by a number of software packages: \texttt{NumPy}, \texttt{SciPy} \citep{2020SciPy-NMeth}, \texttt{Astropy} \citep{2018AJ....156..123T}, \texttt{Matplotlib} \citep{2007CSE.....9...90H}, and \texttt{IPython}/\texttt{Jupyter} \citep{2007CSE.....9c..21P}.
We use \texttt{MultiNest} \citep{2008MNRAS.384..449F,2009MNRAS.398.1601F,2019OJAp....2E..10F} and \texttt{emcee} \citep{2013PASP..125..306F} for sampling posterior distributions, and \texttt{ChainConsumer} \citep{2016JOSS....1...45H} for analysis and plotting.

The authors acknowledge support from the European Research Council in the form of a Consolidator Grant with number 681431.
IH acknowledges support from the Beecroft Trust.

\bibliographystyle{mnras}
\bibliography{main}

\end{document}